\documentclass[twoside,fleqn]{article}
\usepackage{espcrc2}
\usepackage{epsfig}
\def\lsim{\raise0.3ex\hbox{$<$\kern-0.75em\raise-1.1ex\hbox{$\sim$}}}
\def\gsim{\raise0.3ex\hbox{$>$\kern-0.75em\raise-1.1ex\hbox{$\sim$}}}
\newcommand{\ie}{{\sl i.e.~\/}}

\newcommand{\AmS}{{\protect\the\textfont2
  A\kern-.1667em\lower.5ex\hbox{M}\kern-.125emS}}

\title{
\vskip -100pt
\mbox{} \hfill BI-TP 99/31\\
\mbox{} \hfill September 1999\\
\vskip 45pt
Lattice QCD at Finite Temperature and Density}

\author{Frithjof Karsch
\thanks{The work has been supported by the TMR network ERBFMRX-CT-970122
and the DFG under grant Ka 1198/4-1 .} 
\\
\vskip 6pt
Fakult\"at f\"ur Physik, Universit\"at Bielefeld, D-33615 Bielefeld, Germany
}      
\begin{document}

\begin{abstract}
We review recent results on QCD at finite temperature and non-vanishing
baryon number density. We focus on observables which are of immediate
interest to experimental searches for the Quark Gluon Plasma, \ie
the phase transition temperature, the equation of state in two and 
three flavour QCD and thermal effects on hadron masses. 
Some new developments in finite density QCD are also presented. 
\end{abstract}

\maketitle

\section{Introduction}

\subsection{Preface}

There are many reasons to study a complicated interacting field theory 
like QCD under extreme conditions, eg. at finite temperature and/or 
non-vanishing baryon number density. 
Of course, first of all we learn about the collective behaviour of 
strongly interacting matter, its 
critical behaviour, equation of state and thermal properties
of hadrons as well as their fate at high temperature. 
However, in doing so we also have a look at 
the complicated non-perturbative structure of the QCD vacuum from
a different perspective and can learn about the mechanisms behind 
confinement and chiral symmetry breaking. This does allow to 
check concepts that have been developed to
explain these non-perturbative features of QCD. 

\begin{figure}[htb]
%\vskip -0.8truecm
%\vspace{9pt}
\hspace*{-0.4cm}\epsfig{bbllx=7,bblly=50,bburx=530,bbury=450,
file=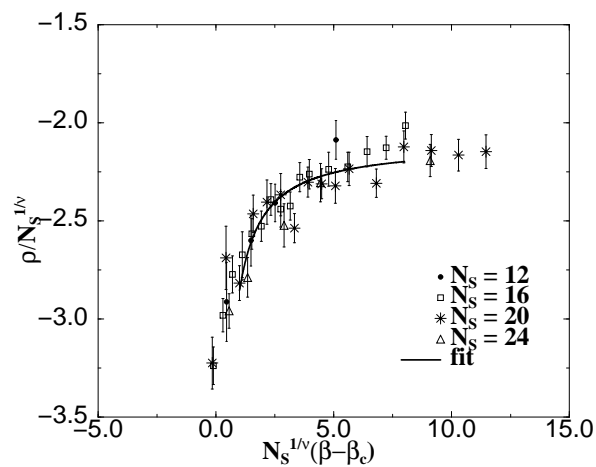,width=78mm}
\vskip -0.7truecm
\caption{Finite size scaling analysis of an order parameter for monopole 
condensation at the deconfinement transition temperature of the SU(2) 
gauge theory \cite{DiG99a,DiG99}.}
\vskip -0.7truecm
\label{fig:monopole}
\end{figure}

This review focuses on the former aspect of finite temperature QCD 
studies; the latter has recently been
discussed in \cite{DiG99a}. In fact, an important step
in the direction of understanding confinement in terms of the
dual superconductor picture has been taken recently by the Pisa
group \cite{DiG99}. They analysed the scaling behaviour of an order 
parameter for monopole condensation and could show that it scales in 
SU(2) and SU(3) gauge theories with the critical exponents expected 
for the deconfinement transition  (Fig.~\ref{fig:monopole}). This is a 
strong hint for the survival of the monopole condensation mechanism 
as a source for confinement in the continuum limit.

\subsection{Outline}

Significant progress in finite temperature lattice QCD is based
on the use of improved actions. In the pure gauge sector we have 
seen that these actions strongly reduce finite cut-off effects
and allow to calculate thermal quantities on rather coarse lattices
in good agreement with continuum extrapolated results. A  similar
approach is now followed in calculations with light quarks. The use of 
improved staggered and Wilson actions leads to new estimates
of the critical temperature and the equation of state which we will
discuss in the following sections. Moreover, thermodynamic
calculations with domain wall fermions (DWF), which can greatly improve 
the chiral properties of lattice fermion actions, reached a stage
where first quantitative results can be reported on $T_c$ and 
thermal masses. The latter will be discussed in Section 4. 
In Section 5 we will discuss new developments in the analysis of
QCD at finite density. Section 6 contains an outlook.

An important part of finite temperature ($T$) and density ($n_B$)
(chemical potential ($\mu$)) calculations consists of trying to understand 
the QCD phase diagram, \ie the order of the phase transition as function 
of the number of flavours ($n_f$) and quark masses ($m_q$) and its 
dependence on $T$ and $n_B$ (or $\mu$). We will not discuss these
issues here. The finite temperature phase diagram has been 
discussed extensively in recent reviews \cite{Uka98} and the
possibly rather complex phase structure of QCD at high 
density and low temperatures has been the topic of  E. Shuryak 
contribution to this conference \cite{Shu99}.

\section{The Critical Temperature} 

\begin{figure*}[htb]
\vskip -0.8truecm
\vspace{9pt}
\hspace*{-0.2cm}\epsfig{file=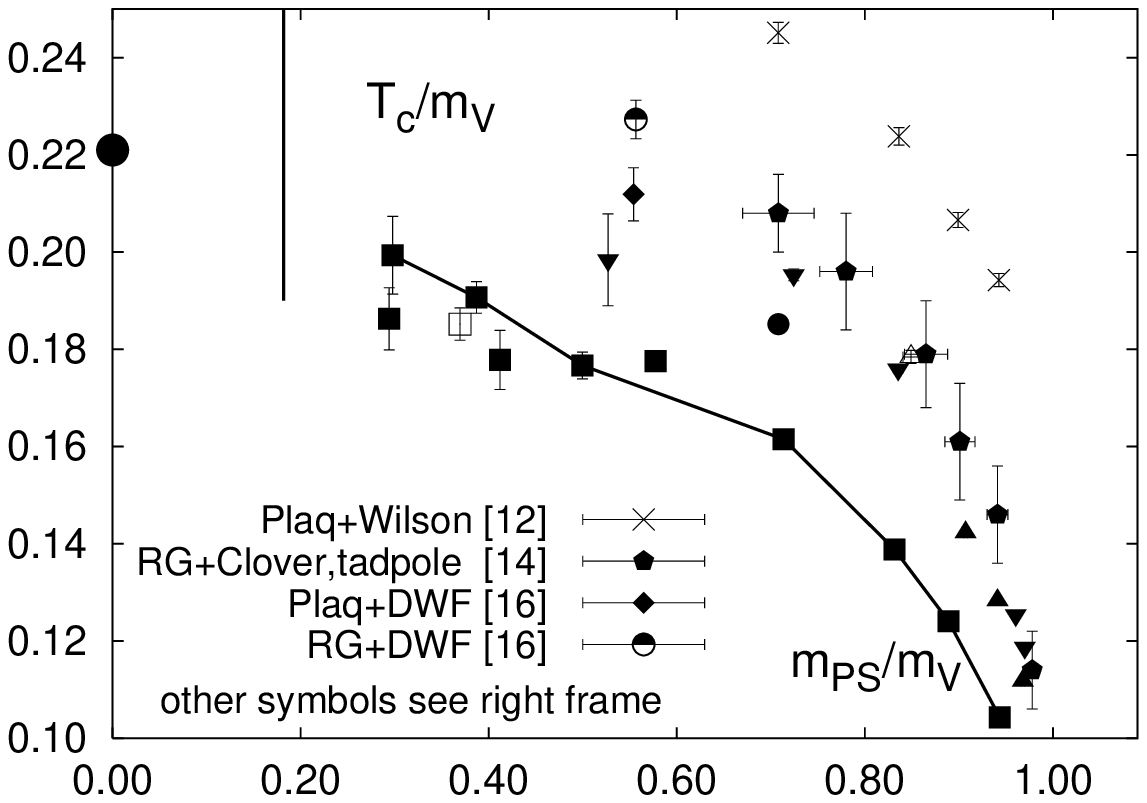,width=82mm}
%\vskip 0.1truecm
\hspace*{-0.2cm}\epsfig{file=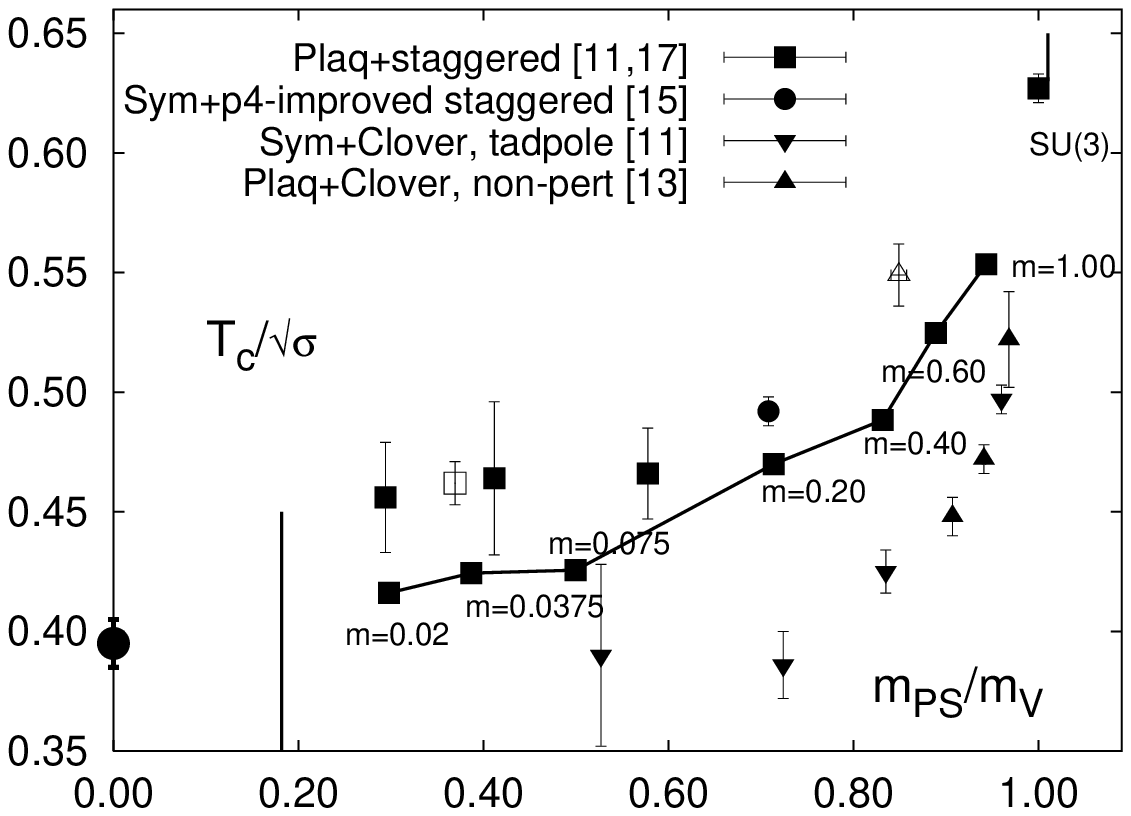,width=82mm}
\vskip -0.7truecm
\caption{
Transition temperatures for 2-flavour QCD in units of $m_V$ (left) 
and $\sqrt{\sigma}$ (right), respectively. The corresponding quark masses 
for staggered fermions are indicated in the right frame.
Filled (open) symbols are for results from $N_\tau = 4$ (6) lattices.
The large dots drawn for $m_{PS}/m_V =0$ are just indicative 
for the chiral limit. 
They would correspond to a critical temperature of $T_c = 170~{\rm MeV}$. 
The long vertical lines indicate the location of the physical limit, 
$m_{PS}\equiv m_\pi = 140~{\rm MeV}$. 
The solid lines connect a data set obtained with the standard staggered
fermion action \cite{BieU}. They are meant to guide the eye.}
%\vskip -0.7truecm
\label{fig:tcnf2}
\end{figure*}

One of the basic goals of lattice QCD calculations at finite T
is to provide quantitative results for the phase transition temperature.
In the pure gauge sector this goal has been achieved. The value of the
critical temperature for the deconfining phase transition in a $SU(3)$
gauge theory is known with small statistical errors. Remaining systematic 
errors are of the order of 3\%. They arise from the calculation of the 
string tension at $T=0$ which is used to set the scale for $T_c$.
In Table 1 we summarise results for $T_c/\sqrt{\sigma}$ obtained
with different gauge actions on lattices with varying temporal extent
$N_\tau \sim 1/(aT)$ and extrapolated to the continuum limit ($N_\tau 
\rightarrow \infty$).  

\begin{table}[b]
% space before first and after last column: 1.5pc
% space between columns: 3.0pc (twice the above)
%\setlength{\tabcolsep}{1.5pc}
% -----------------------------------------------------
% adapted from TeX book, p. 241
%\newlength{\digitwidth} \settowidth{\digitwidth}{\rm 0}
\catcode`?=\active \def?{\kern\digitwidth}
% -----------------------------------------------------
\vspace*{-0.5cm}
\caption{Critical temperature of the $SU(3)$ gauge theory in units 
of $\sqrt{\sigma}$ 
obtained from calculations with different gauge actions and
extrapolated to the continuum limit.}
\label{tab:tc}
\begin{center}
\vspace{0.3cm}
\begin{tabular}{|l|l|l|}
\hline
action & Ref.& $T_c/\sqrt{\sigma} $\\ \hline
standard Wilson & [5,6] &$ 0.630~(5)$\\
Symanzik impr. (tree level) & [6] &$ 0.634~(4)$\\
RG-improved & [7] &$ 0.650~(5)$\\
\hline
\end{tabular}
\end{center}
\end{table}

Using for the string tension the value\footnote{This 
may be deduced from quenched spectrum calculations 
($m_\rho /\sqrt{\sigma} = 1.81~(4)$ \cite{Wit97}). Recent estimates
from CP-PACS data seem to lead to a somewhat smaller value \cite{Akira}.}
$\sqrt{\sigma} \simeq 425~{\rm MeV}$, we find  
$T_c \simeq 270~{\rm MeV}$. This large value can be understood
in terms of the particle spectrum of the quenched theory; in the low
temperature, confining phase there exist only rather heavy glueballs 
($m_G \gsim 1.7$~GeV).
A rather large temperature thus is needed to create a sufficiently 
dense glueball gas, which can trigger a deconfining transition.  

Investigations of the critical temperature in QCD with quarks of 
finite mass indeed have shown that the transition temperature drops 
rapidly with
decreasing quark masses\footnote{At intermediate values of $m_q$ 
the transition only reflects a rapid cross-over in thermodynamic
observables rather than a true phase transition.
The peak in the chiral susceptibility or
the Polyakov-Loop susceptibility defines in this case a 
{\it pseudo-critical} temperature. We will continue to call this 
temperature the {\it transition} temperature at finite values of $m_q$ 
For 2-flavour QCD it is expected to be a {\it critical}
temperature, corresponding to a phase transition, only in the chiral 
limit. For 3-flavour QCD it is a critical temperature below a 
certain critical quark mass \cite{Aok99,Pei99}.}.

Unlike in the pure gauge theory the transition temperature for QCD with
finite quark masses does seem to be strongly affected by the 
discretization scheme used for the fermion action. The early 
calculations with the standard staggered \cite{Ber97}
and Wilson \cite{Bit91} actions led to widely different estimates for $T_c$.
Finite-cut off effects are thus expected to be large
and improvement of the fermion action should be expected to be 
important. Indeed, the new calculations performed with improved 
Wilson (Clover) \cite{Ber97,Edw99,Eji99} and staggered \cite{Pei99b} 
actions as  well as with DWF \cite{Vra99} yield transition 
temperatures which are in much better agreement among each other. 
Unfortunately, this statement is, at present, only partially correct. 
In fact, for comparable values of $m_q$, \ie fixed ratios of pseudo-scalar
(``pion'') and vector meson (''rho'') masses, $m_{PS}/m_{V}$, 
it only holds when we set the scale for $T_c$ using a hadron mass, 
eg. $m_V$. When we follow a similar 
approach and use the string tension, $\sqrt{\sigma}$, to set the scale 
the agreement is less evident. In this case calculations with the Wilson 
fermion
action typically lead to transition temperatures about 20\% below the 
results obtained with staggered fermion actions. 
The current status of the determination of $T_c$ 
for 2-flavour QCD is summarised in Fig.~\ref{fig:tcnf2}.

As can be seen the simulations with Clover fermions 
performed by different groups are consistent with each other.
Results on $T_c/m_V$ do not seem to depend significantly on the gauge 
action (one plaquette Wilson \cite{Edw99}, Symanzik improved \cite{Ber97} 
or RG-improved \cite{Eji99}) nor 
does it seem to be important whether tadpole \cite{Ber97,Eji99} or 
non-perturbative \cite{Edw99} Clover coefficients are used. Also the DWF
results \cite{Vra99} seem to be 
insensitive to the gauge action chosen (plaquette or RG-improved) and are 
consistent
with results obtained with the Clover action. Results obtained
with an improved staggered fermion action \cite{Pei99b}, 
the p4-action\footnote{The
p4-action improves the rotational symmetry of the quark propagator in
${\cal O}(p^4)$. It strongly reduces the cut-off effects in
bulk thermodynamic observables \cite{Hel99}, see Fig.~4.}, 
also agree with these data within 10\%. 

Unfortunately this consistent picture is, at present,
not reproduced when calculating $T_c$ in units of $\sqrt{\sigma}$ 
(right frame of Fig.~\ref{fig:tcnf2}). A possible source for this 
discrepancy might be
the calculation of the heavy quark potential, which in the
case of Wilson fermions so far has only been performed on rather small
spatial lattices, e.g. $8^3\times 16$. This may lead to an overestimate
of the string tension. It, however, also is possible that calculations
on $N_\tau =4$ lattices are still performed at too strong coupling
and do not allow for a unique determination of the scale. A hint
in this direction may be the large difference in $T_c/\sqrt{\sigma}$
observed in calculations on $N_\tau =4$ and 6 lattices 
with the non-perturbatively improved Clover action \cite{Edw99}. 
Clearly more work is needed here to establish a unique result
for $T_c$ using different fermion formulations.

\begin{figure}[htb]
%\vskip -0.8truecm
%\vspace{9pt}
\hspace*{-0.2cm}\epsfig{file=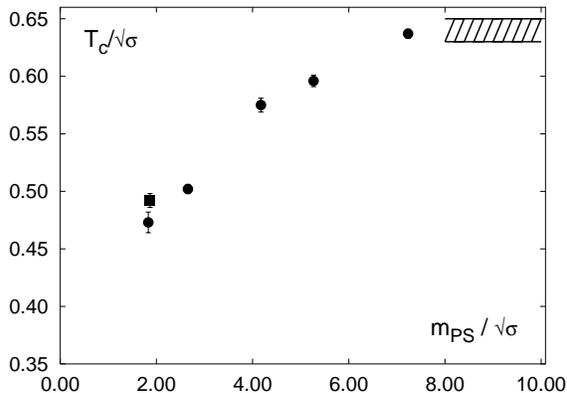,width=80mm}
\vskip -0.7truecm
\caption{The transition temperature in 2 (square) and 3 (circles)
flavour  QCD versus $m_{PS}/\sqrt{\sigma}$
using the p4-action. The dashed band indicates
the uncertainty on $T_c/\sqrt{\sigma}$ in the quenched limit.}
\vskip -0.5truecm
\label{fig:tc_pion}
\end{figure}

In general, we note that the transition temperature obtained with improved
actions tends to be larger than what previously has been quoted on the
basis of calculations performed with the standard staggered 
action. If the quark mass dependence does not change drastically closer 
to the 
chiral limit\footnote{Calculations within the framework of quark-meson
models suggest a rapid drop of $T_c$ for $m_{PS}< m_\pi$ \cite{Wet99}.} 
the current data suggest
\begin{equation}
T_c \simeq (170-190)~{\rm MeV}
\end{equation}
for 2-flavour QCD in the chiral limit.
In fact, this estimate also holds for 3-flavour QCD. Calculations 
which are currently performed for $n_f=2$ and 3 using the same
improved staggered fermion action, suggest
that the flavour dependence of $T_c$ is rather weak \cite{Pei99b} 
(see Fig.~\ref{fig:tc_pion}).

It is remarkable that the transition temperature drops 
significantly already in a region where all hadron masses are quite large.
This is apparent from Fig.~\ref{fig:tc_pion} where we show
$T_c$ in units of $\sqrt{\sigma}$ plotted 
vs. $m_{PS} / \sqrt{\sigma}$ for $n_f=2$ and 3. 
As can be seen the transition
temperature starts deviating from the quenched values for $m_{PS}\lsim
(6-7)\sqrt{\sigma}\simeq 2.5~{\rm GeV}$. We also note that the 
dependence of $T_c$ on $m_{PS}/\sqrt{\sigma}$
is almost linear in the entire mass interval. This might be expected
for light quarks in the vicinity of 
a $2^{nd}$ order chiral transition where the pseudo
critical temperature depends on the mass of the Goldstone-particle
like
\begin{equation}
T_c(m_{\pi}) - T_c(0) \sim m_{\pi}^{1/\beta\delta} ~.
\end{equation} 
For 2-flavour QCD where the critical indices are expected to 
belong to the universality class of 3-d, $O(4)$
symmetric spin models one would indeed expect $1/\beta\delta=1.1$. 
However, this clearly cannot be the origin of the quasi linear
behaviour observed for large hadron masses independent\footnote{A
similar conclusion holds for $n_f=2$, when one analyses the unimproved
standard staggered fermion data.} of $n_f$. A resonance gas model
would probably be more appropriate to describe the thermodynamics
for these heavy quarks.  

\section{The Equation of State}

\begin{figure*}[htb]
\vskip -0.8truecm
\vspace{9pt}
\hspace*{-0.2cm}\epsfig{file=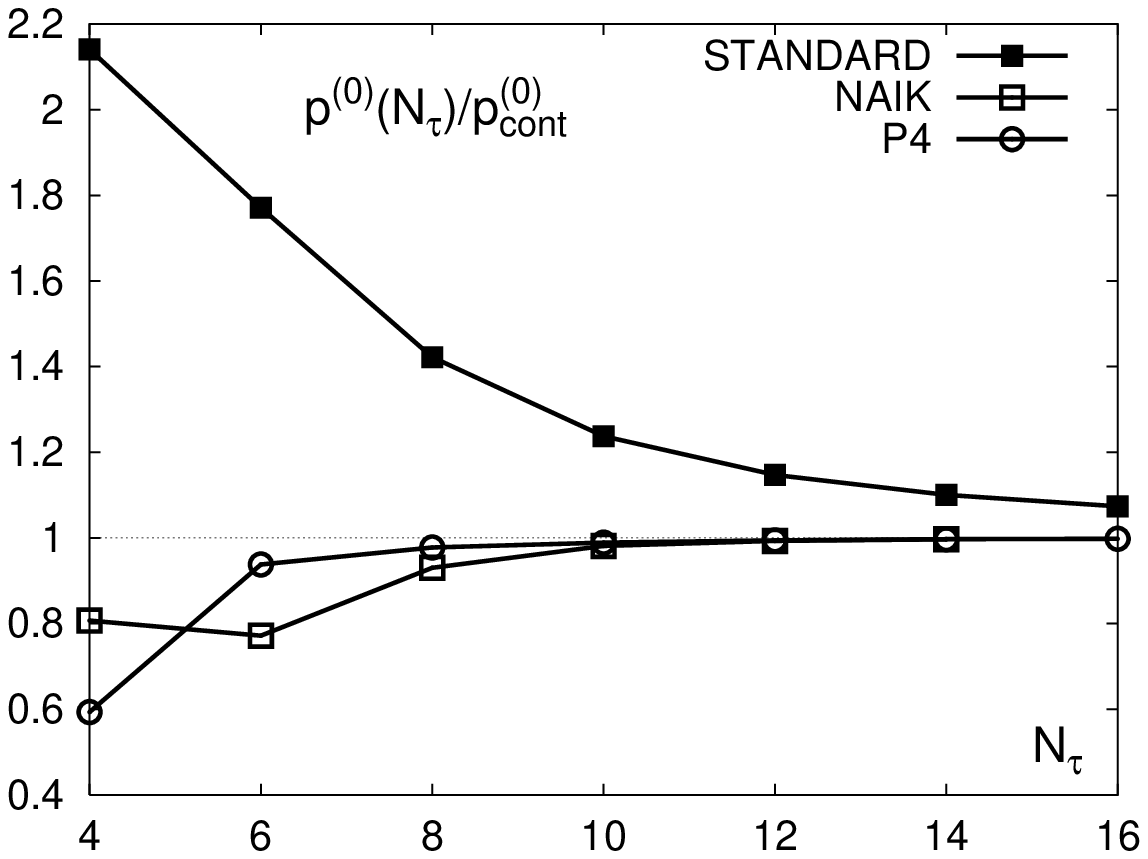,width=82mm}
%\vskip 0.1truecm
\hspace*{-0.2cm}\epsfig{file=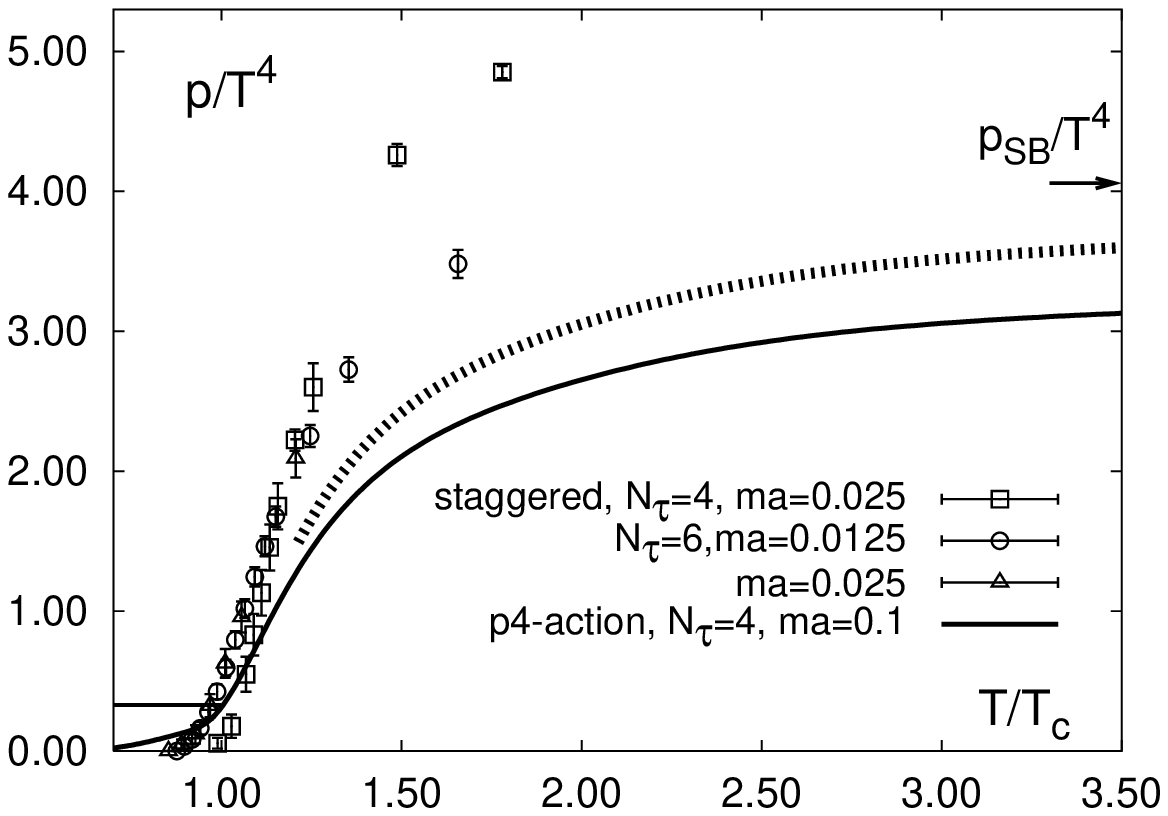,width=82mm}
\vskip -0.7truecm
\caption{Cut-off dependence of the pressure of an ideal Fermi gas 
calculated on spatially infinite lattices with temporal extent $N_\tau$ 
and different staggered fermion actions (left) and the pressure for 
2-flavour QCD calculated by using the standard staggered fermion action
\cite{Ber97b} and the p4-improved action \cite{Pei99b} (right).
The dashed band shows an estimate for the 2-flavour pressure in the 
continuum limit (see text). The horizontal line for $T\le T_c$ shows the 
pressure of a massless pion gas.} 
%\vskip -0.7truecm
\label{fig:pressure_nf2}
\end{figure*}

In the pure gauge sector bulk thermodynamic quantities such as the
pressure and energy density have been analysed in detail. It has
been verified that the most significant cut-off effects result from
high momentum modes, which dominate the infinite temperature, ideal gas 
limit. Improved actions that lead to small cut-off effects in the 
ideal gas limit\footnote{In the ideal gas limit 
the cut-off dependence
can be analysed analytically \cite{Eng82,Bei96}. For a recent analysis
of staggered fermion actions see \cite{Hel99}.} still do so 
at finite temperature \cite{Bei99}. 

In a recent
analysis the CP-PACS collaboration calculated the pressure and
energy density using the RG-improved (Iwasaki) action \cite{Oka99}. 
They confirm that after extrapolation to the continuum limit also this 
action yields results for the $SU(3)$ equation of state which are,
within an error of (3-4)\%,
consistent with the continuum extrapolation obtained from the Wilson
\cite{Boy96}, improved \cite{Bei99} as well as some
fixed point actions \cite{Pap96}. In fact,
the Wilson and RG-improved actions represent extreme cases for such 
calculations.   Their cut-off corrections have opposite sign which  
on coarse ($N_\tau=4$) lattices leads to deviations of the
pressure by about 25\% from the continuum extrapolated result.
In view of this the agreement reached with different discretization 
schemes is rather reassuring.    

Also in the presence of light quarks the use of an improved action
thus seems to be mandatory, if one wants to calculate bulk thermodynamic
quantities. The standard staggered and Wilson actions are known
to lead to large deviations from the continuum ideal gas behaviour
on coarse lattices (small $N_\tau$) \cite{Eng82}. For staggered
fermions improved actions, leading to smaller cut-off effects in 
thermodynamic observables, can be constructed by adding suitably
chosen three-link terms to the conventional one-link 
terms \cite{Hel99,Nai89}. In Fig.~\ref{fig:pressure_nf2}
(left frame) we show the cut-off dependence of the ideal gas 
pressure obtained from these minimally improved staggered fermion actions 
(Naik \cite{Nai89} and p4 \cite{Hel99} action)\footnote{Staggered actions 
with even smaller cut-off effects have been constructed \cite{P6,Bie97}.
However, they also require a significantly larger numerical effort.}.

As expected the systematics seen for the cut-off dependence in the ideal
gas limit carries over to finite T. In the right frame of 
Fig.~\ref{fig:pressure_nf2} we show results for the pressure of 2-flavour
QCD obtained with the standard staggered action (and Wilson gauge action) 
on lattices with temporal extent $N_\tau=4$ and 6 \cite{Ber97b}. These
are compared with results obtained with the p4-action (and a Symanzik
improved gauge action) \cite{Pei99,Hel99}. In the latter case also 
fat links \cite{Blu97} have been introduced in the one-link terms to 
improve the flavour symmetry of the staggered action.

While the pressure calculated with the standard staggered action
rapidly overshoots the ideal gas limit (as expected from the analysis
of the cut-off effects in the ideal gas limit) the results
obtained with the p4-action stay below the ideal gas limit and
show a temperature dependence very similar to what has been found 
in the pure gauge sector. 
We note that the calculations with the p4-action
have been performed with rather large quark masses, $m/T=0.4$,
corresponding to $m_{PS}/m_{V} \simeq 0.7$, while the 
calculation with standard staggered action are for $m_{PS}/m_{V} \simeq
0.3$. In the high temperature phase this does not seem to constitute
a major problem\footnote{Calculations for 4-flavour QCD with 
quark masses $m/T=0.2$ and $0.4$ show no significant quark mass
dependence \cite{Jos97}. Moreover, also in the continuum the pressure of 
an ideal gas of fermions with mass $m/T= 0.4$ deviates by
less than 10\% from that of a massless gas.}. Smaller quark masses
are, however, definitely needed close to $T_c$ and below in order to 
become sensitive to the contributions from light pion modes. 

In the case of the SU(3) gauge theory the magnitude of
the cut-off dependence in the temperature range up to a few times $T_c$
has been found to be about half of what has been calculated for the ideal 
gas, infinite temperature limit. If this continues to hold for the fermion 
sector one should expect, that the current result, obtained with  
the p4-action on $N_\tau=4$ lattices, underestimates the continuum  
result by about 15\% for $T\gsim T_c$. Based on this consideration an 
estimate for the continuum extrapolated pressure is also shown in 
Fig.~\ref{fig:pressure_nf2}.  We also note that a calculation performed
with identical, improved staggered fermion actions shows that the pressure 
normalised to the 
continuum ideal gas value has the same temperature dependence in 2 and
3 flavour QCD \cite{Pei99b}.
 
\begin{figure}[htb]
%\vskip -0.8truecm
%\vspace{9pt}
\hspace*{-0.5cm}\begin{picture}(8.5,6.6)
%\hspace*{-0.2cm}
\put(0.2,0.0){\epsfig{file=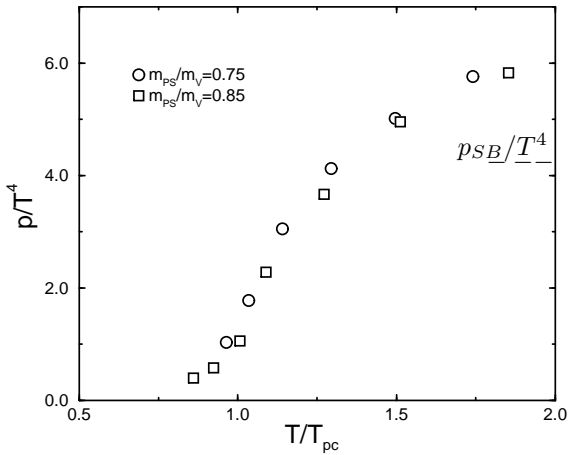,width=85mm}}
\put(6.6,4.4){$p_{SB}/T^4$}
\put(6.9,4.2){${\bf ---}$}
\end{picture}
\vskip -1.1truecm
\caption{Pressure vs. $T/T_c$ calculated with Wilson fermions
on a $16^3\times 4$ lattice for two different 
meson mass ratios, $m_{PS}/m_V$.} 
\vskip -0.5truecm
\label{fig:wilson_pressure}
\end{figure}

Unfortunately, Wilson actions with similarly good high temperature
behaviour have not been constructed so far. The Clover action does
not improve the ideal gas behaviour, \ie it has the same infinite
temperature limit as the Wilson action. 
%Furthermore, the need to
%adjust the hopping parameter appropriately makes calculations of bulk
%thermodynamic quantities more demanding than in the case of staggered 
%fermions. 
Nonetheless, a first attempt to calculate the equation of state, using the 
tadpole-improved Clover action combined with the RG-improved gauge action, 
has now been undertaken by the CP-PACS collaboration \cite{Eji99}.
A preliminary result for the pressure in 2-flavour QCD is shown
in Fig.~\ref{fig:wilson_pressure}. Similar to simulations with the 
standard staggered fermion action
one observes an overshooting of the ideal gas limit
reflecting the cut-off effects in the unimproved fermion sector.

\section{Thermal Masses}

Understanding the temperature dependence of hadron properties,
e.g. their masses and widths, is of central importance for the
interpretation of heavy ion experiments. Thermal modifications
of the heavy quark potential influence the spectrum of heavy quark
bound states.
Their experimentally observed suppression \cite{jpsi} thus is expected to
be closely linked to the deconfining properties of QCD above $T_c$ 
\cite{Satz}. Changes in the chiral condensate, on the
other hand, influence the light hadron spectrum and may leave 
experimental signatures, for instance in the enhanced 
dilepton production observed in heavy ion experiments \cite{lepton}.

In numerical calculations on Euclidean lattices one has
access to thermal Green's functions $G_H(\tau, \vec{r})$
in fixed quantum number channels, $H$, to which in particular at high 
temperature many excited states contribute. As the temporal
direction of the Euclidean lattice is rather short at finite 
temperature one usually has not enough information on the correlation
function to extract reliably thermal effects on the (pole) masses.  
A way out may come from the use of anisotropic lattices, which has
extensively been explored by the QCDTARO group \cite{For99b}. In their
recent analysis with quenched Wilson fermions they show evidence for
a rapid change in the thermal correlation functions across $T_c$ which
indicates a rapid change in thermal masses above $T_c$. At the same time
their analysis of pseudo-scalar wavefunctions does, however, show
evidence that the correlation among quarks in this quantum number channel 
remains strong. In how far this hints at the presence of 
light ''pionic'' bound states has to be analysed further. 

Anisotropic lattices in combination with NRQCD have also been
used to analyse thermal effects on heavy quark bound states \cite{Fin98}.
Large mass shifts have been observed for the first excited states.

The missing information on the long distance behaviour of 
thermal correlation functions may also be overcome by using refined
techniques to analyse the numerical data on thermal correlation functions.
It recently has been suggested that the maximum entropy method \cite{Jar96}
may help to extract more detailed information on thermal modifications
of hadronic spectral functions \cite{Asa99}.

\begin{figure}[htb]
%\vskip -0.8truecm
\vspace{9pt}
\hspace*{-0.2cm}\epsfig{file=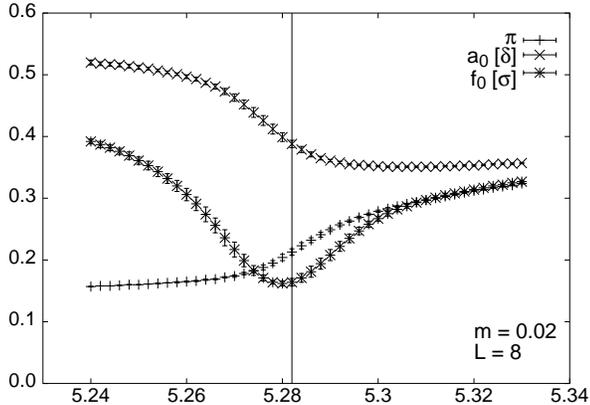,width=78mm}
%\vskip 0.1truecm
\vskip -0.7truecm
\caption{Susceptibilities in different quantum number channels for
2-flavour QCD with staggered fermions on lattices of size $8^3\times 4$. 
The various points are obtained from a Ferrenberg-Swendsen reweighting 
analysis based on measurements at a few $\beta$-values in the interval 
shown. The horizontal line indicates the location of $\beta_c$.}
\vskip -0.7truecm
\label{fig:masses}
\end{figure}

So far the most convincing evidence for modifications of hadron properties
at high temperature comes from the analysis of the behaviour of correlation
function at large spatial separations, which yields screening masses 
\cite{DeTar}, as  well as thermal susceptibilities, 
\begin{equation}
\chi_H = \int_0^{1/T} {\rm d}\tau \int {\rm d}^3 r\; G_H(\tau, \vec{r})~,
\end{equation}
with $G_H$ denoting the hadron correlation
function in the quantum number channel $H$. If there is only a single
stable particle of mass $m_H$ contributing to 
$G_H$ then $\chi_H \sim m_H^{-2}$. In general the susceptibilities,
however, define only {\it effective masses}, \ie they average over
contributions from the ground state and all excited states
in a particular quantum number channel. 

In Fig.~\ref{fig:masses} we show $1/\sqrt{\chi_H}$ for 2-flavour 
QCD obtained with staggered fermions of mass $m_q=0.02$ on lattices of
size $8^3\times 4$.. This figure is based on data from \cite{Kar94}.
As can be seen the $f_0$ and $\pi$ 
masses become (almost) degenerate at $\beta_c$ while 
the $a_0$ remains heavy at $\beta_c$.
The former behaviour is expected, $f_0$ and $\pi$ correlation functions
are related through a rotation in $SU(2)$ flavour space. The 
degeneracy thus reflects the restoration of the $SU(2)$ flavour symmetry. 
The difference of $\chi_{a_0}$ and $\chi_\pi$ on the other hand
reflects the persistence of the $U_A(1)$ symmetry breaking. 
A crucial question 
here is how the gap observed for non-zero $m_q$ changes in the
chiral limit. At high
temperature topologically non-trivial gauge field configurations are 
expected to be suppressed, which in turn would lead to a strong 
reduction in the strength of $U_A(1)$ symmetry breaking and thus in a 
strong reduction of the mass splitting between $a_0$ and $\pi$.
For 2-flavour QCD it is expected that the quark mass dependence 
is quadratic, $(m_{a_0} - m_\pi) \sim (\chi_\pi - \chi_{a_0}) \sim
A + B m_q^2$. Previous investigations of this were, however, not 
conclusive \cite{Uka98}. If a quadratic ansatz is assumed calculations
with staggered fermions led to the conclusion that a non-zero
mass splitting remains also above $T_c$, \ie the $U_A(1)$ remains
broken \cite{Uka98}. The problem has now been addressed again by
the Columbia group \cite{Vra99} using DWF. Due 
to the improved chiral properties of this action one should 
find a quadratic dependence on the quark mass. For $T \simeq 1.2 T_c$
the Columbia group indeed does observe such a quark mass dependence.
In the chiral limit they find a non-zero mass splitting 
from the susceptibilities as well as from the analysis of screening
masses \cite{Vra99},
\begin{equation}
m_{a_0} - m_\pi = 0.0606~(67) + 9.66~(58) m_q^2 ~.
\end{equation} 
The $U_A(1)$ thus remains broken above $T_c$, although the mass splitting 
is strongly reduced (see Fig.~\ref{fig:masses}). This
picture is also supported by an analysis of the disconnected
parts of flavour singlet correlation functions 
in quenched QCD using DWF \cite{Sin99}. A vanishing of these
would signal a degeneracy between the $\sigma$ and $\eta$ mesons. 
The disconnected parts get non-zero contributions 
only from topologically non-trivial configurations \cite{Kog98}.
Indeed, these are strongly suppressed above $T_c$. However, at 
$T=1.25 T_c$
about 10\% of the configurations do still carry a non-zero 
topological charge indicating the persistence of $U_A(1)$ 
symmetry breaking above $T_c$.

\section{Finite Density QCD}

\begin{figure*}[htb]
\vskip -0.8truecm
\vspace{9pt}
\hspace*{-0.2cm}\epsfig{file=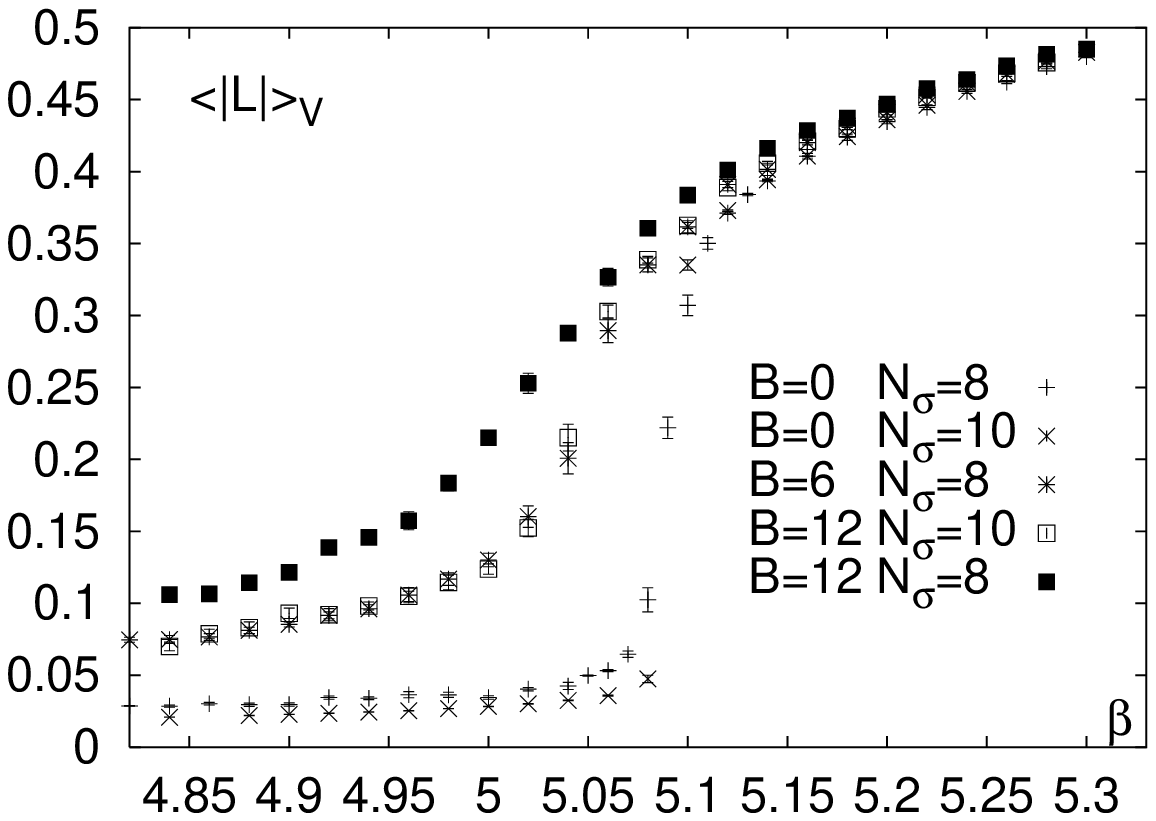,width=82mm}
%\vskip 0.1truecm
\hspace*{-0.2cm}\epsfig{file=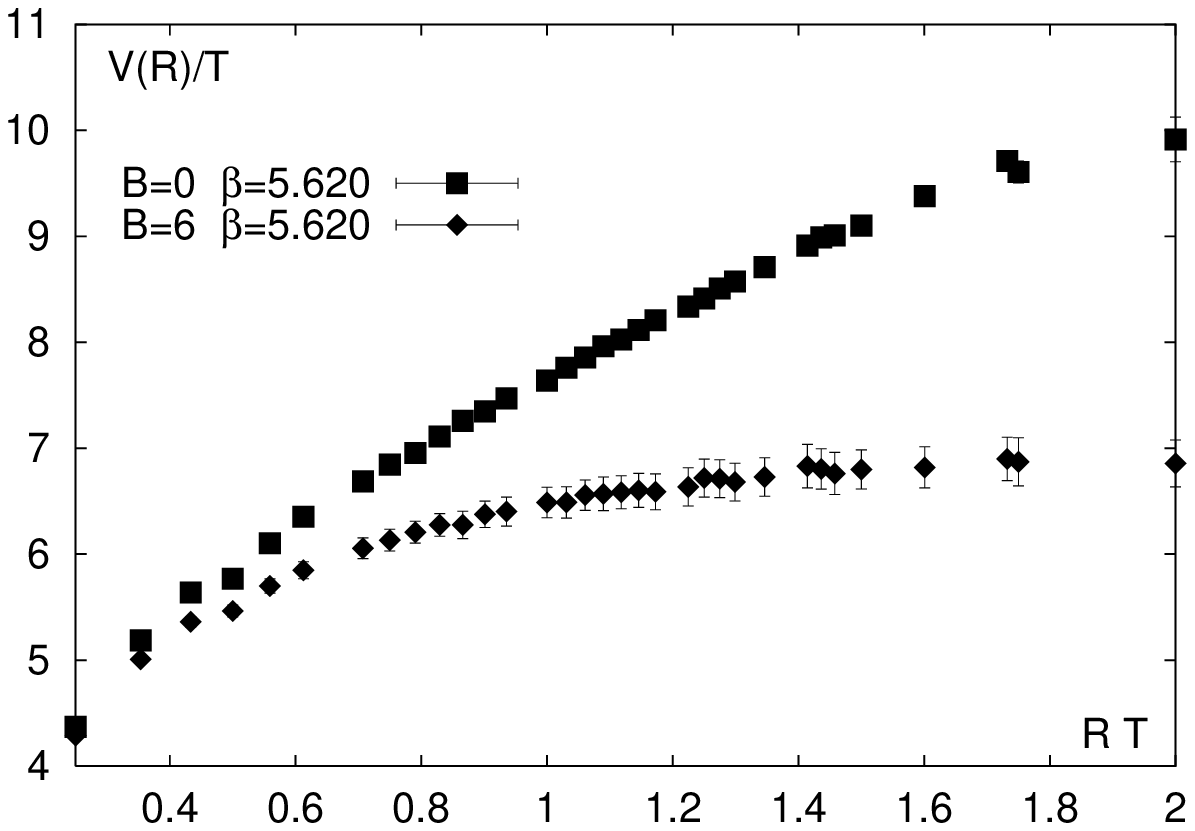,width=82mm}
\vskip -0.7truecm
\caption{Polyakov loop expectation value (left) calculated on
$N_\sigma^3 \times 2$ and the heavy quark
potential (right) calculated on $16^3\times 4$ lattices in quenched QCD 
at zero and non-zero baryon number, $B/3$.}
\vskip -0.3truecm
\label{fig:density}
\end{figure*}

Finite density calculations in QCD are affected by the well known
sign problem, \ie the fermion determinant becomes complex for
non-zero values of the chemical potential $\mu$ and thus prohibits the
use of conventional numerical algorithms. The most detailed studies
have been performed so far using the Glasgow algorithm \cite{Bar98}, 
which is based on a fugacity expansion of the grand canonical partition 
function at non-zero $\mu$,

\begin{equation}
Z_{GC}(\mu/T,T,V) = \sum_{B=-\alpha V}^{\alpha V} z^B Z_B(T,V) ~,
\label{gcp}
\end{equation}
where $z=\exp{(\mu /T)}$ is the fugacity and $Z_B$ are the canonical
partition functions for fixed quark number $B$; $\alpha = 3,6$ for
one species of staggered or Wilson fermions, respectively. However, this 
approach so far did not overcome the severe numerical difficulties.

It thus seems necessary to approach the finite density problems
from another point of view. A reformulation of the original ansatz
may lead to a representation of the partition function which, in
the ideal case, would require the averaging over configurations
with strictly positive weights only, or at least would lead to a
strong reduction of  configurations with negative weights. 

A big step away from the original formulation is to start
from a Hamiltonian approach \cite{Wie99a}. Here it has been shown 
that problems with a fluctuating integrand can successfully be 
reformulated in terms of a model where the configurations generated do
have strictly positive weights (meron cluster algorithm \cite{Wie99b}).
Whether such an approach can be applied to QCD remains to be seen.
 
An alternative formulation of finite density QCD is given in terms
of canonical rather than grand canonical partition functions \cite{Mil87},
\ie rather than introducing a non-zero chemical potential through
which the number density is controlled one introduces directly a
non-zero baryon number (or quark number $B$) from which the 
baryon number density on lattices of size $N_\sigma^3 \times N_\tau$
is obtained as
${n_B / T^3} = {B\over  3} ({N_\tau / N_\sigma})^3$.

After introducing a complex chemical potential in $Z_{GC}$(Eq.~\ref{gcp})
the canonical partition functions can be obtained via a Fourier 
transform\footnote{The use of this ansatz for the calculation of canonical 
partition functions as expansion coefficients for $Z_{GC}$ has been 
discussed in \cite{Has92,Alf99}.},
\begin{equation}
Z_B(T,V) = {1\over 2\pi} \int_0^{2\pi}{\rm d}\phi \; {\rm e}^{i\phi B/T}\;
Z_{GC} (i\phi,T,V)~ .
\end{equation}
Also this formulation is by no means easy to use in general, \ie
for QCD with light quarks. In particular, it also still suffers
from a sign problem. It, however, leads to a quite natural and 
useful formulation of the quenched limit of QCD at non-zero
density \cite{Eng99b}. 

%
%\subsection{The quenched limit of QCD at finite density}
\subsection{Quenched limit of finite density QCD}

It had been noticed early that the 
straightforward replacement of the fermion determinant by a constant
does not lead to a meaningful static limit of QCD \cite{Bar86}. In 
fact, this simple replacement corresponds to the static limit 
%($m_q\rightarrow \infty$) 
of another 
theory with an equal number of fermion flavours carrying baryon number $B$
and $-B$, respectively \cite{Ste96}. This should not be
too surprising. When one starts with 
%QCD at 
%a non-zero baryon number
and takes the limit of infinitely heavy quarks something should be 
left over from the determinant that represents the objects that carry 
the baryon number. In the canonical formulation this becomes obvious. 
For $m_q\rightarrow \infty$ one ends up with a partition function, which 
for baryon number $B/3$ still includes the sum over products of $B$ 
Polyakov loops, \ie the static quark propagators which carry 
the baryon number \cite{Eng99b}. This limit also has some
analogy in the grand canonical formulation where the coupled limit
$m_q,\mu\rightarrow \infty$ with $\exp{(\mu)}/2m_q$
kept fixed has been performed \cite{Sta91,Blu96}\footnote{This is a
well known limit in statistical physics. When deriving the 
non-relativistic gas limit from a relativistic gas of particles
with mass $\bar{m}$, the rest mass is splitted off from the chemical 
potential, $\mu \equiv \mu_{nr} + \bar{m}$,
in order to cancel the corresponding rest mass term in the particle
energies. On the lattice $\bar{m} = \ln (2m_q)$ for large bare quark
masses.}.  

In the confined phase of QCD the baryon number is 
carried by the rather heavy nucleons. Approximating them by 
static objects may thus be quite reasonable and we may expect to
get valuable insight into the thermodynamics of QCD at non-zero 
baryon number density already from quenched QCD.

From a numerical point of view there is hope that we can get control
over this limit using different approaches. At non-zero $\mu$ 
a variant of the Glasgow approach \cite{Alo99} seems to
become applicable for large quark masses and also the static limit
of Blum et al. \cite{Blu96} may be explored further \cite{For99}.

In the canonical approach simulations at non-zero $B$ 
can be performed on relatively large lattices and the use of baryon number 
densities up to a few times nuclear matter density may be possible
\cite{Eng99b,Kac99}.
The simulations performed so far show the basic features expected at 
non-zero density. As can be seen from the behaviour of the Polyakov loop 
expectation value shown in Fig.~\ref{fig:density}
the transition region gets shifted to smaller temperatures
(smaller coupling $\beta$). The broadening of the transition region may
suggest a smooth crossover behaviour at
non-zero density. However, in a canonical simulation it also may indicate 
the presence of a region of coexisting phases and thus would signal 
the existence of a $1^{st}$ order phase transition. This deserves further 
analysis. 

Even more interesting is the behaviour of the heavy quark 
potential in the low temperature phase. As shown in the right frame of 
Fig.~\ref{fig:density} the potential does get screened at  
non-vanishing number density. This will have a direct influence
on heavy quark bound states at high density.

\section{Outlook}

We have focused in this review on the calculation of basic thermodynamic 
quantities which are of immediate interest to experimental searches for 
the Quark Gluon Plasma. 

In quenched QCD the critical temperature and the equation of state
have been calculated on the lattice and extrapolated to the continuum
with an accuracy of a few percent. These calculations set a 
benchmark for many analytical studies of QCD thermodynamics \cite{And99}.
The progress made in developing and testing
improved fermion actions for thermodynamic calculations shows that 
a similar accuracy for QCD with light quarks is within reach.
The current systematic studies with different improved fermion actions
may soon lead to a determination of the transition temperature and the 
equation of state with similar accuracy.

We do have reached some understanding of thermal effects on hadron
properties. In particular, modifications of the light meson spectrum
due to flavour and approximate $U_A(1)$ symmetry restoration have
been established. However, lattice calculations~so~far~did~not~come~up
with detailed quantitative results on thermal masses, which could be 
confronted with 
%or used to analyse 
experimental data. 
There are a few promising
ans\"atze which can lead to more detailed information on thermal
modifications of hadronic spectral densities. 

Of course, there are many more important issues which have to be
addressed in the future. Even at vanishing baryon number density
we do not yet have a satisfactory understanding of the critical
behaviour of 2-flavour QCD in the chiral limit and 
the physically realized situation of QCD with two light, nearly
massless quarks and a heavier strange quark has barely been
analysed.     

Moreover, the entire phase diagram at non-zero baryon number 
density is largely unexplored. An interesting phase structure
is predicted at high density and low temperatures which currently
is not accessible to lattice calculations. 
%Here we probably should
%first stick to more moderate goals and try to get control over finite 
%density QCD with heavy quarks.
This does require new algorithmic developments.

There are thus many interesting questions 
%left unanswered which await
waiting
to be answered in the next millennium.

\medskip
\noindent
{\bf Acknowledgements:} I would like to thank the CCP 
at the University  of Tsukuba for its kind hospitality during the
time this talk has been prepared and written up. I also want to thank
N. Christ, S. Ejiri, M. Okamoto, D.K. Sinclair, I.O. Stamatescu,
D. Toussaint, and U.-J. Wiese for communication on their results and 
K. Kanaya and A. Ukawa for comments on the manuscript.

\end{document}